\documentclass[twocolumn]{article}
\pdfoutput=1
\usepackage{amsmath}
\usepackage{amssymb}
\usepackage{amsfonts}
\usepackage[affil-it]{authblk}
\usepackage{xfrac}
\newcommand{\ket}[1]{\left| #1 \right>} 
\newcommand{\braket}[2]{\left< #1 \vphantom{#2} \right| \left. #2 \vphantom{#1} \right>} 
\newcommand{\ketbra}[2]{|#1\rangle\langle#2|}

\begin{document}

\title{An introduction to quantum machine learning}

\author{Maria Schuld$^{a}$, Ilya Sinayskiy$^{a,b}$ and Francesco Petruccione$^{a,b}$}
\affil{$^{a}${\em{Quantum Research Group, School of Chemistry and Physics,  University of KwaZulu-Natal, Durban, KwaZulu-Natal, 4001, South Africa}}\\
$^{b}${\em{National Institute for Theoretical Physics (NITheP), KwaZulu-Natal, 4001, South Africa}}}\vspace{6pt}
\date{\today}

\twocolumn[
  \begin{@twocolumnfalse}
    \maketitle
    \begin{abstract}
    Machine learning algorithms learn a desired input-output relation from examples in order to interpret new inputs. This is important for tasks such as image and speech recognition or strategy optimisation, with growing applications in the IT industry. In the last couple of years, researchers investigated if quantum computing can help to improve classical machine learning algorithms. Ideas range from running computationally costly algorithms or their subroutines efficiently on a quantum computer to the translation of stochastic methods into the language of quantum theory. This contribution gives a systematic overview of the emerging field of quantum machine learning. It presents the approaches as well as technical details in an accessable way, and discusses the potential of a future theory of quantum learning.\\


    \end{abstract}
  \end{@twocolumnfalse}
]

\vspace{1cm}

\section{Introduction}
Machine learning refers to an area of computer science in which patterns are derived (`learned') from data with the goal to make sense of previously unknown inputs. As part of both artificial intelligence and statistics, machine learning algorithms process large amounts of  information for tasks that come naturally to the human brain, such as image and speech recognition, pattern identification or strategy optimisation. These problems gain significant importance in our digital age, an illustrative example being Google's PageRank machine learning algorithm for search engines that was patented by Larry Page in 1997\footnote{See https://www.princeton.edu/~achaney/tmve/wiki100k/ docs/PageRank.html [Last accessed 6/24/2014]} and led to the rise of what is today one of the biggest IT companies in the world. Other important applications of machine learning are spam mail filters, iris recognition for security systems, the evaluation of consumer behaviour, assessing risks in the financial sector or developing strategies for computer games. In short, machine learning comes into play wherever we need computers to interpret data based on experience. This usually involves huge amounts of previously collected input-output data pairs, and machine learning algorithms have to be very efficient in order to deal with so called \textit{big data}. \\

Since the volume of globally stored data is growing by around 20\% every year (currently ranging in the order of several hundred exabytes \cite{hilbert11}), the pressure to find innovative approaches to machine learning is rising. A promising idea that is currently investigated by academia as well  as in the research labs of leading IT companies exploits the potential of quantum computing in order to optimise classical machine learning algorithms. In the last decades, physicists already demonstrated the impressive power of quantum systems for information processing. In contrast to conventional computers built on the physical implementation of the two states `0' and `1', quantum computers can make use of a qubit's superposition of two quantum states $\ket{0}$ and $\ket{1}$ (e.g. encoded in two distinct energy levels of an atom) in order to follow many different paths of computation at the same time. But the laws of quantum mechanics also restrict our access to information stored in quantum systems, and coming up with quantum algorithms that outperform their classical counterparts is very difficult. However, the toolbox of quantum algorithms is by now fairly established and contains a number of impressive examples that speed up the best known classical methods \cite{nielsen10}. The technological implementation of quantum computing is emerging \cite{georgescu14}, and many believe that it is only a matter of time until the numerous theoretical proposals can be tested on real machines. On this background, the new research field of quantum machine learning might offer the potential to revolutionise future ways of intelligent data processing.\\

A number of recent academic contributions explore the idea of using the advantages of quantum computing in order to improve machine learning algorithms. For example, some effort has been put into the development of quantum versions  \cite{rigatos07, behrman13, gupta01} of artificial neural networks (which are widely used in machine learning), but they are often based on a more biological perspective and a major breakthrough has not been accomplished yet \cite{schuld14b}. Some authors try to develop entire quantum algorithms that solve problems of pattern recognition \cite{ventura00, trugenberger02, schutzhold03}. Other proposals suggest to simply run subroutines of classical machine learning algorithms on a quantum computer, hoping to gain a speed up \cite{lloyd13, rebentrost13, wiebe14}. An interesting approach is adiabatic quantum machine learning, which seems especially fit for some classes of optimisation problems \cite{neven09, pudenz13, neigovzen09}. Stochastic models such as Bayesian decision theory or hidden Markov models find an elegant translation into the language of open quantum systems \cite{sentis12, clark14}. Despite this growing level of interest in the field, a comprehensive theory of quantum learning, or how quantum information can in principle be applied to intelligent forms of computing, is only in the very first stages of development.\\

\begin{figure}[t]
  \centering    \includegraphics[width=0.5\textwidth]{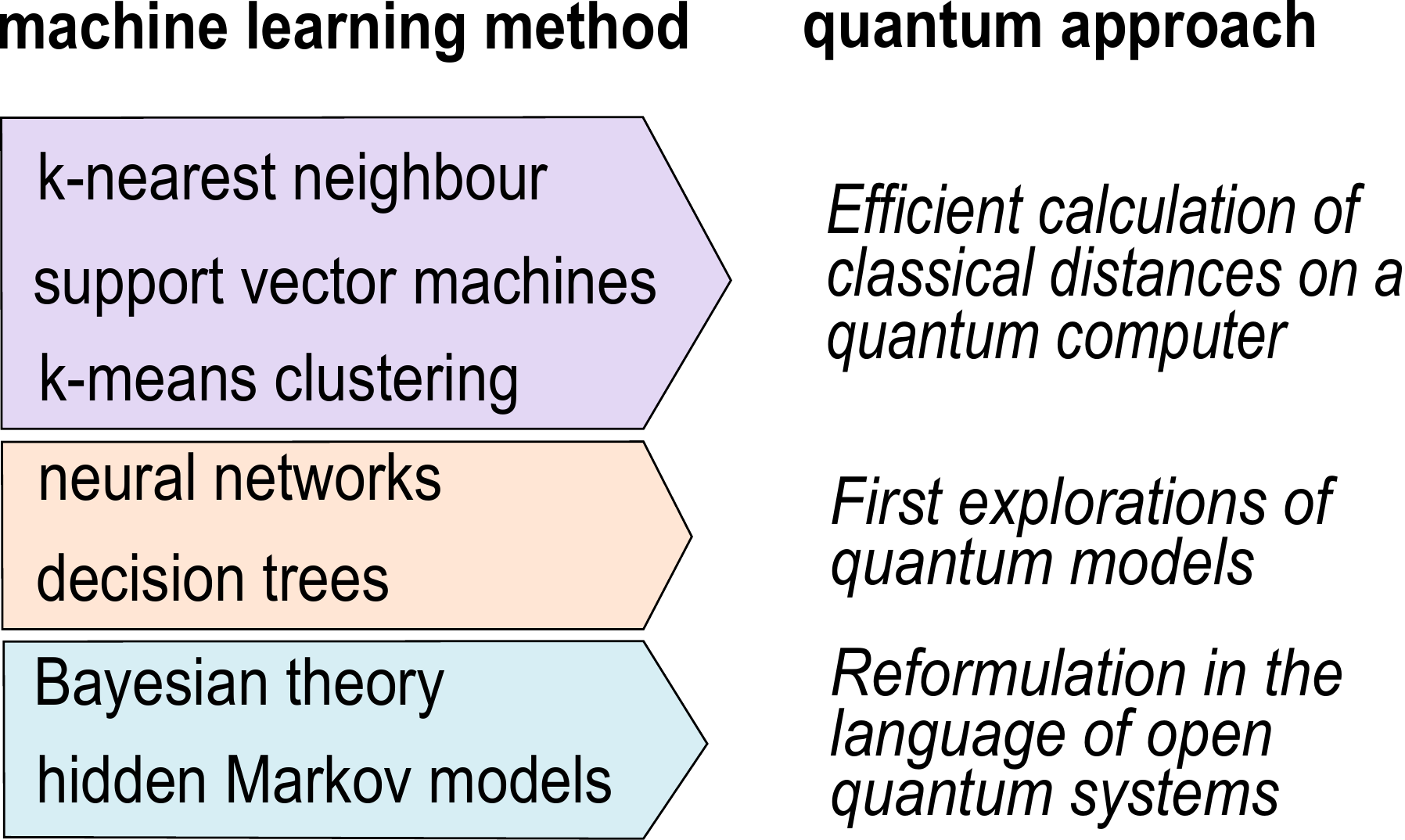}
 \caption{Overview of methods in machine learning and approaches from a quantum information perspective as presented in this paper. }
 \label{figure1}
\end{figure}

This contribution gives a systematic overview of the emerging field of quantum machine learning, with a focus on methods for pattern classification. After a brief discussion of the concepts of classical and quantum learning in Section \ref{learning}, the paper is divided into seven sections, each presenting a standard method of machine learning (namely $k$-nearest neighbour methods, support vector machines, $k$-means clustering, neural networks, decision trees, Bayesian theory and hidden Markov models) and the various approaches to relate each method to quantum physics. This structure mirrors the still rather fragmented field and allows the reader to select specific areas of interest. As summarised in Figure \ref{figure1}, for $k$-nearest neighbour methods, support vector machines and $k$-means clustering, authors are mainly concerned to find efficient calculations of classical distances on a potential quantum computer, while probabilistic methods such as Bayesian theory and hidden Markov models find an analogy in the formalism of open quantum systems. Neural networks and decision trees are still waiting for a convincing quantum version, although especially the former has been a relatively active field of research in the last decade. Finally, in Section \ref{conc} we briefly discuss the need for future works on quantum machine learning that concentrate on how the actual \textit{learning} part of machine learning methods can be improved using the power of quantum information processing.

\section{Classical and quantum learning}\label{learning}

\subsection{Classical machine learning}

The theory of machine learning is an important subdiscipline of both artificial intelligence and statistics, and its roots can be traced back to the beginnings of artificial neural network and artificial intelligence research in the 1950's \cite{russell10, rosenblatt58}. In 1959, Arthur Samuel gave his famous definition of machine learning as the `field of study that gives computers the ability to learn without being explicitly programmed'\footnote{It is interesting to note that although quoted in numerous introductions to machine learning, the original reference to the machine learning pioneer's most famous statement is very difficult to find. Authors either refer to other secondary publications, or falsely cite Samuel's seminal paper from 1959 \cite{samuel00}.}. This is in fact misleading, since the algorithm itself does not adapt in the learning process, but the function it encodes. In more formal language, this means that the input-output relation of a computer program is derived from a set of training data (which is often very big). Such methods gain importance as computers increasingly interact with humans and have to become more flexible to adapt to our specific needs. A prominent example is a spam mail filter that learns from user behaviour and external databases to classify new spam mails correctly. However, this is only one of many different cases where machine learning intersects with our every-day lives. \\

\begin{figure}[t]
 \centering    \includegraphics[width=0.35\textwidth]{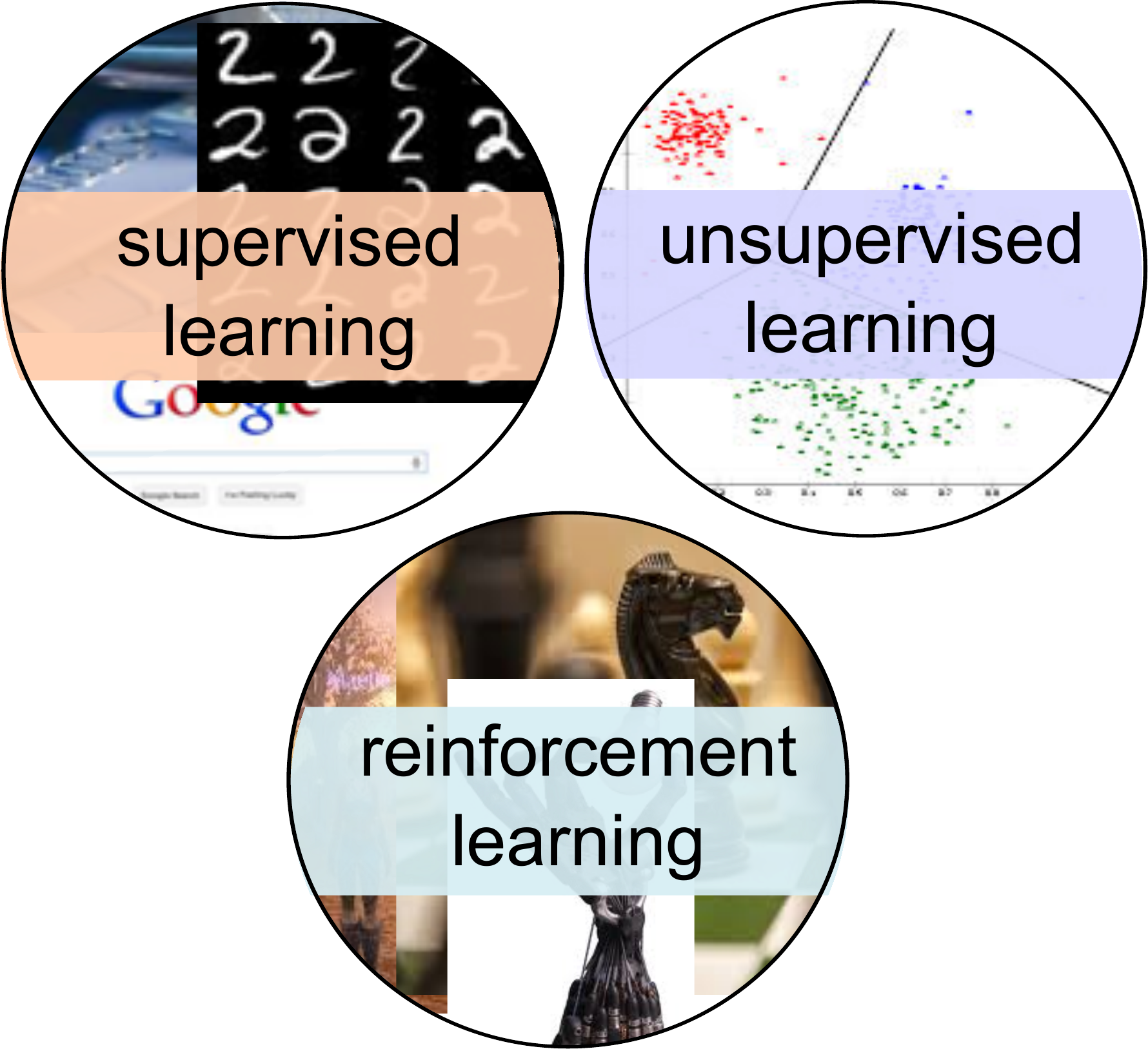} 
\caption{The three types of classical learning. Supervised learning derives patterns from training data and finds application in pattern recognition tasks. Unsupervised learning infers information from the structure of the input and is important for data clustering. Reinforcement learning optimises a strategy due to feedback by a reward function, and usually applies to intelligent agents and games.  }
 \label{figure2}
\end{figure} 

In the theory of machine learning, the term \textit{learning} is usually divided into three types (see Figure \ref{figure2}), which help to illustrate the spectrum of the field: \textit{supervised}, \textit{unsupervised} and \textit{reinforcement learning}. In supervised learning, a computer is given examples of correct input-output relations and has to infer a mapping therefrom. Probably the most important task is \textit{pattern classification}, where vectors of input data have to be assigned to different classes. This might sound like a rather technical problem, but is in fact something humans do continuously - for example when we recognise a face from different angles and light conditions as belonging to one and the same person, or when we classify signals from our sensory organs as dangerous or not. We could even go so far and say that pattern classification is the abstract description of `interpreting' input coming from our senses. It is no surprise that a big share of machine learning research tries to imitate this remarkable ability of human beings with computers, and there is an entire zoo of algorithms that generalise from large training data sets how to classify new input.\\

The second category, \textit{unsupervised learning}, has not been part of machine learning for a long time, as it describes the process of finding patterns in data without prior experience or examples. A prominent task is data clustering, or forming subgroups out of a given dataset, in order to summarize large amounts of information by only a few stereotypes. This is for example an important problem in sociological studies and market research. Note that this task is closely related to classification, since clustering means effectively to assign a class to each vector of a given set, but without the goal of treating new inputs. \\

Finally, \textit{reinforcement learning} is the closest to what we might associate with the expression `learning'. Given a framework of rules and goals, an agent (usually a computer program that acts as a player in a game) gets rewarded or punished depending on which strategy it uses in order to win. Each reward reinforces the current strategy, while punishment leads to an adaptation of its policy \cite{alpaydin04,duda12}. Reinforcement learning is a central mechanism in the development and study of intelligent agents. However, it will not be in the focus of this paper, and it differs in many regards from the other two types of learning. Investigations into quantum games and quantum intelligent agents are diverse and numerous (see for example, \cite{landsburg11,eisert99,briegel12,du02,piotrowski03}), and shall be treated elsewhere. \\

Even within these categories, the expression `learning' can relate to different procedures. For example, it may refer to a training phase in which optimal parameters of an algorithm (e.g. weights, initial states) are obtained. This is done by presenting examples of correct input-output-relations to a task, and adapting the parameters to reproduce these examples. The training set is then discarded \cite{bishop06}. An illustrative case close to human learning is the weight adjustment process in artificial neural networks through backpropagation or deep learning \cite{hinton06,rumelhart88}. Training phases are often the most costly part of a machine learning algorithm and efficient training methods become especially important when dealing with so called \textit{big data}. Besides learning as a parameter optimisation problem, there is a large number of machine learning algorithms that do not have an explicit learning phase. For example, if presented with an unclassified input vector, the $k$-nearest-neighbour for pattern classification uses the training data to decide upon its classification. In this case, learning is not a parameter optimisation problem, but rather a decision function inferred from examples. In reinforcement learning, this decision function becomes a full \textit{strategy}, and learning refers to the adaptation of the strategy to increase the chances of future reward.  \\

Whatever type and procedure of learning is chosen, optimal machine learning algorithms run with minimum resources and have a minimum error rate related to the task (as indicated by misclassification of input, poor division into clusters, little reward of a strategy). Challenges lie in the problem of finding parameters and initial values that lead to an optimal solution, or to come up with schemes that reduce the complexity class of the algorithm.\footnote{The complexity of a problem tells us by what factor the computational resources needed to solve a problem grow if we increase the input to the problem (e.g. the digits of a number) by one.} This is where quantum computing promises to help. 

\subsection{Quantum machine learning}

Quantum computing refers to the manipulation of quantum systems in order to process information. The ability of quantum states to be in a superposition can thereby lead to a substantial speedup of a computation in terms of complexity, since operations can be executed on many states at the same time. The basic unit of quantum computation is the qubit,  $\ket{\psi} = \alpha \ket{0} + \beta \ket{1}$ (with $\alpha,\beta \in \mathbb{C}$ and $\ket{0}, \ket{1}$ in the two-dimensional Hilbert space $\mathcal{H}^2$). The absolute squares of the amplitudes are the probability to measure the qubit in the $0$ or the $1$ state, and quantum dynamics always maintain the property of probability conservation given by $|\alpha|^2 + |\beta|^2 =1$. In mathematical language this means that transformations that map quantum states onto other quantum states (so called \textit{quantum gates}) have to be unitary. Through single qubit quantum gates we are able to manipulate the basis state, amplitude or phase of a qubit (for example through the so called X-gate, the Z-gate and the Y-gate respectively), or put a qubit with $\beta=0$ ($\alpha =0$) into an equal superposition $\alpha=\beta=\sfrac{1}{\sqrt{2}}$ ($\alpha=\sfrac{1}{\sqrt{2}}, \beta= -\sfrac{1}{\sqrt{2}}$) (the Hadamard or H-gate). Multi-qubit gates are often based on controlled operations that execute a single qubit operation only if another (ancilla or control qubit) is in a certain state. One of the most important gates is the two qubit XOR-gate, which flips the basis state of the second qubit in case the first qubit is in state $\ket{1}$. A two-qubit gate that will be mentioned later is the SWAP-gate exchanging the state of two qubits with each other.\\

\begin{figure}[t]
 \centering    \includegraphics[width=0.3\textwidth]{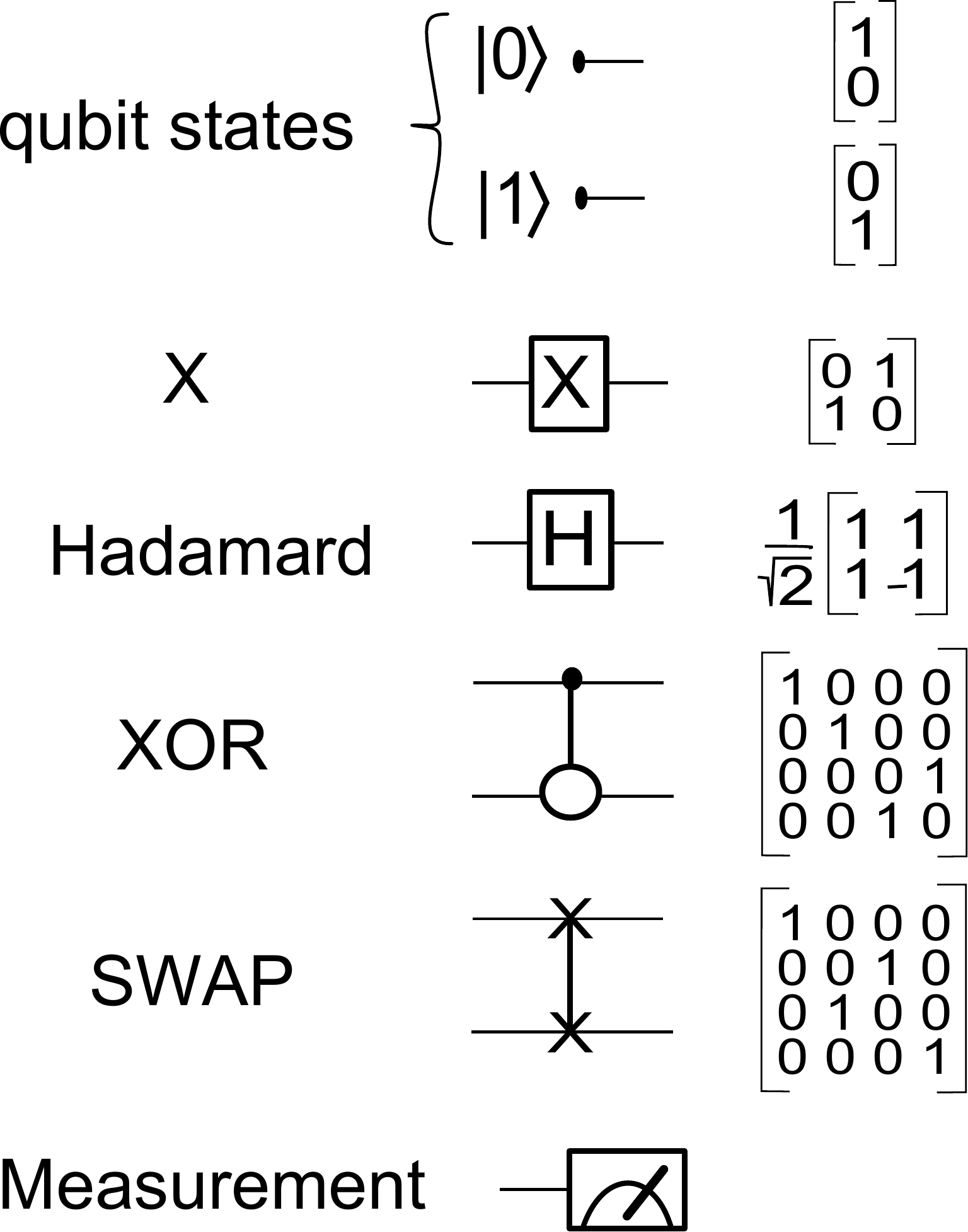} 
\caption{Representation of qubit states, unitary gates and measurements in the quantum circuit model and in the matrix formalism.}
 \label{figure3}
\end{figure} 

Quantum gates are usually expressed as unitary matrices (see also Figure \ref{figure3}). The matrices operate on $2^n$-dimensional vectors that contain the amplitudes of the $2^n$ basis states of a $n$-dimensional quantum system. For example, the XOR-gate working on the quantum state $\ket{\psi} = \sfrac{1}{\sqrt{2}} \; (\ket{00}+\ket{11})$ would look like
\[ \begin{pmatrix} 1 & 0 & 0 & 0\\  0 & 1 & 0 & 0\\ 0 & 0 & 0 & 1\\ 0 & 0 & 1 & 0 \end{pmatrix}\; \cdot \;  \frac{1}{\sqrt{2}}\begin{pmatrix} 1\\0\\0\\1 \end{pmatrix} = \frac{1}{\sqrt{2}}\begin{pmatrix} 1\\0\\1\\0 \end{pmatrix}, \]
and produce $ \ket{\psi'} = \sfrac{1}{\sqrt{2}}\; (\ket{00}+\ket{10})$. The art of developing algorithms for a potential quantum computer is to use such elementary gates in order to create a quantum state that has a relatively high amplitude for states that represent solutions for the given problem. A measurement in the computational basis then produces such a desired result with a relatively high probability. Quantum algorithms are usually repeated a number of times since the result is always probabilistic. For a comprehensive introduction into quantum computing, we refer to the standard textbook by Nielsen and Chuang \cite{nielsen10}.\\

In quantum machine learning, quantum algorithms are developed to solve typical problems of machine learning using the efficiency of quantum computing. This is usually done by adapting classical algorithms or their expensive subroutines to run on a potential quantum computer. The expectation is that in the near future, such machines will be commonly available for applications and can help to process the growing amounts of global information. The emerging field also includes approaches vice versa, namely well-established methods of machine learning that can help to extend and improve quantum information theory. \\

As mentioned before, there is no comprehensive theory of quantum learning yet. Discussions of elements of such a theory can be found in \cite{sasaki02, aimeur13, hunziker03}. Following the remarks above, a theory of quantum learning would refer to methods of quantum information processing that learn input-output relations from training input, either for the optimisation of system parameters (for example unitary operators, see \cite{bisio10}) or to find a `quantum decision function' or `quantum strategy'. There are many open questions of how an efficient quantum learning procedure could look like. For example, how can we efficiently implement an optimisation problem (that is usually solved by iterative and dissipative methods such as gradient descent) on a coherent and thus reversible quantum computer? How can we translate and process important structural information, such as distance metrics, using quantum states? How do we formulate a decision strategy in terms of quantum physics? And the overall question, is there a general way how quantum physics can in principle speed up certain problems of machine learning?\\

An underlying question is also the representation of classical data by quantum systems. The most common approach in quantum computing is to represent classical information as binary strings $(x_1,...x_n)$ with $ x_i\in\{0,1\}$ for $i=1,...,n$, that are directly translated into $n$-qubit quantum states $\ket{x_1...x_n}$ from a $2^n$-dimensional Hilbert space with basis $\{\ket{0....00}, \ket{0....01},..., \ket{1....11}\}$, and to read information out through measurements. However, existing machine learning algorithms are often based on an internal structure of this data, for example the Euclidean distance as a similarity measure between two examples of features. Alternative data representations have been proposed by Seth Lloyd and his co-workers, who encode classical information into the norm of a quantum state, $\braket{x}{x} = |\vec{x}|^{-1} \vec{x}^2$, leading to the definition \cite{lloyd13, rebentrost13}
\begin{equation} \ket{x} =|\vec{x}|^{-\sfrac{1}{2}} \vec{x}. \label{lloyddef} \end{equation}
In order to use the strengths of quantum mechanics without being confined by classical ideas of data encoding, finding `genuinely quantum' ways of representing and extracting information could become vital for the future of quantum machine learning.

\section{Quantum versions of machine learning algorithms}

Before proceeding to the discussion of classical machine learning algorithms and their quantum counterparts, we have to take a look on the actual problems these methods intend to solve, as well as introduce the formalism used throughout this article. Probably the most important application is the task of \textit{pattern classification}, and there are many different classical algorithms tackling this problem. Based on a set of training examples consisting of feature vectors\footnote{A feature vector has entries that refer to information on a specific case, in other words a datapoint.} and their respective class attributes, the computer has to correctly classify an unknown feature vector. For example, the feature vector could contain preprocessed information on patients and their correctly diagnosed disease. A machine learning algorithm then has to find the correct disease of a new patient. More precisely, given a training set $\mathcal{T} =\{\vec{v}^p, c^p\}_{p=1,...,N}$ of $N$ $n$-dimensional feature vectors $\vec{v}$ and their respective class $c^p$, as well as a new $n$-dimensional input vector $\vec{x}$, we have to find the class $c^x$ of vector $\vec{x}$. Closely related to pattern classification are other tasks such as \textit{pattern completion} (adding missing information to an incomplete input), \textit{associative memory} (retrieving one of a number of stored memory vectors upon an input) or \textit{pattern recognition} (including finding and examining the shape of patterns; this term is often used as a synonym to pattern classification).  \\

The central problem of unsupervised learning is clustering data. Given a set of feature vectors $\{\vec{v}^p\}$, the goal is to assign each vector to one out of $k$ different clusters so that similar inputs share the same assignment. Other problems of machine learning concern optimal strategies in terms of an unknown reward function, given a set of consecutive observations of choices and consequences. As stated above we will not concentrate on the learning of strategies here. 

\subsection{Quantum versions of $k$-nearest neighbour methods}

\begin{figure}[t]
 \centering    \includegraphics[width=0.2\textwidth]{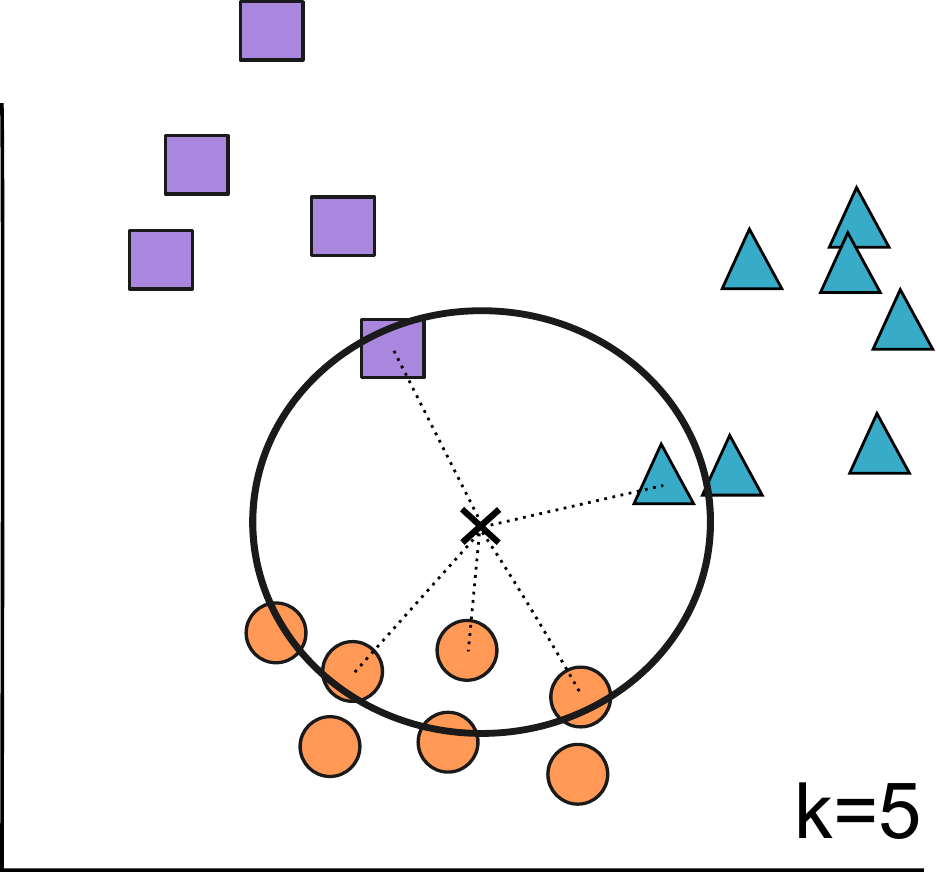}~~~ \includegraphics[width=0.2\textwidth]{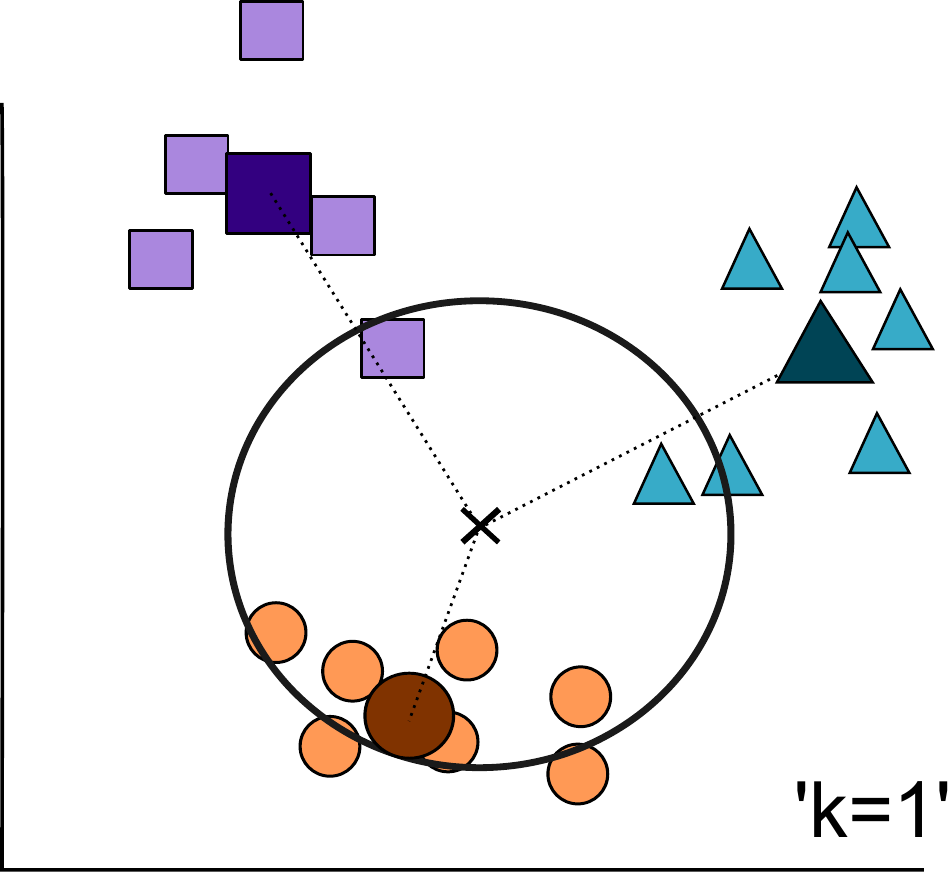}  
\caption{ (Colour online) a:  Illustration of the kNN method of pattern classification. The new vector (black cross) gets assigned to the class that the majority of its $k$ closest neighbours have (in this case it would be the orange circle shape). b: A variation is the nearest-centroid method in which the closest mean vector of a class of vectors defines the classification of a new input. This can be understood as a $k$-nearest neighbour method with preprocessed data and $k=1$.}
 \label{figure4}
\end{figure}

A very popular and simple standard textbook method for pattern classification is the $k$-nearest neighbour algorithm. Given a training set $\mathcal{T}$ of feature vectors with their respective classification as well as an unclassified input vector $\vec{x}$, the idea is to choose the class $c^x$ for the new input that appears most often amongst its $k$ nearest neighbours (see Figure \ref{figure4}). This is based on the assumption that `close' feature vectors encode similar examples, which is true for many applications. Common distance measures are thereby the inner product, the Euclidian or the Hamming distance\footnote{The Hamming distance between two binary strings is the number of flips needed to turn one into the other \cite{hamming50}.}. Choosing $k$ is not always easy and can influence the result significantly. If $k$ is chosen too big we loose the locality information and end up in a simple majority vote over the entire training set, while a very small $k$ leads to noise-biased results. A variation of the algorithm suggests not to run it on the training set, but to calculate the means or centroid $\sfrac{1}{N_c} \sum_p \vec{v}^p$ of all $N_c$ vectors belonging to one class $c$ beforehand, and to select the class of the nearest centroid (we call this here the nearest-centroid algorithm). Another variation weights the influence of the neighbours by distance, gaining an independence of the parameter $k$ (the weighted nearest neighbours algorithm \cite{hechenbichler04}). Methods such as $k$-nearest neighbours are obviously based on a distance metric to evaluate the similarity of two feature vectors. Efforts to translate this algorithm into a quantum version therefore focus on the efficient evaluation of a classical distance through a quantum algorithm. \\

\begin{figure}[t]
 \centering    \includegraphics[width=0.3\textwidth]{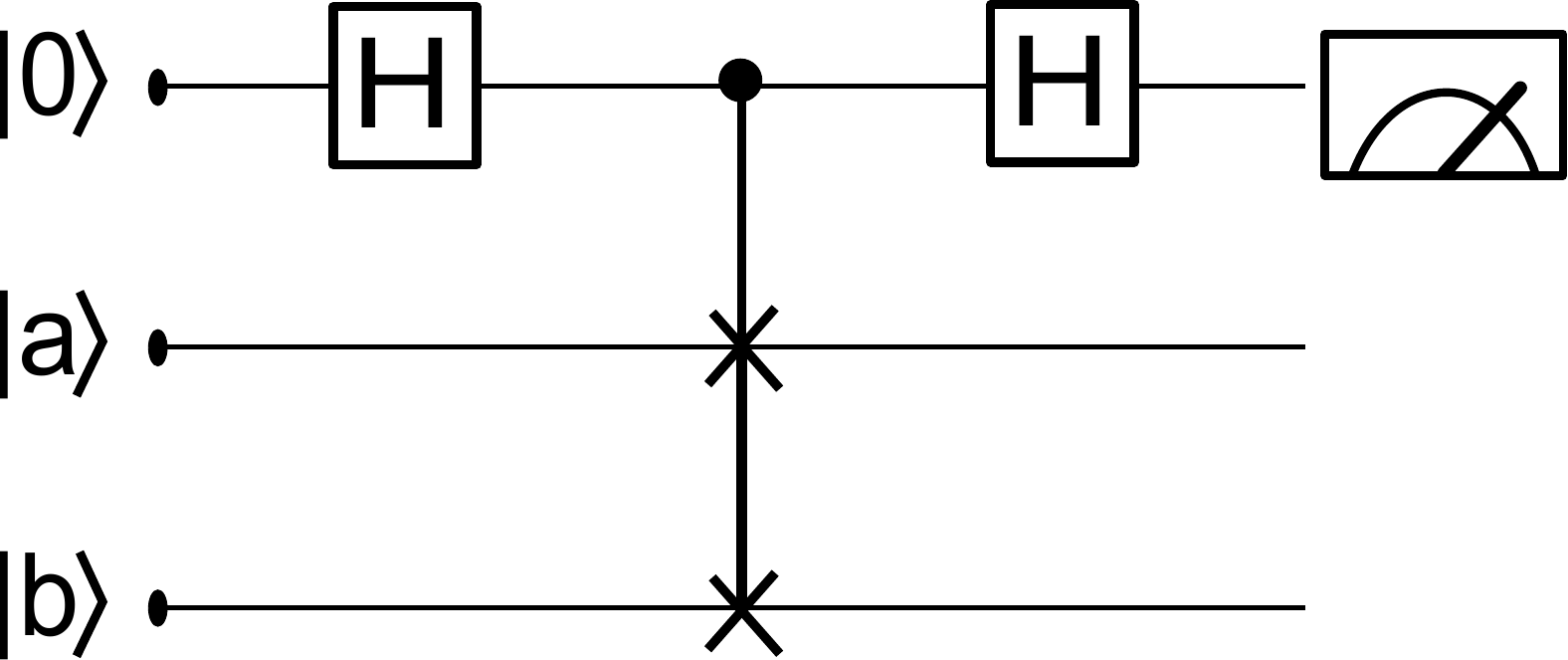} 
\caption{Quantum circuit representation of a swap test routine.}
 \label{figure5}
\end{figure}

A{\"\i}meur, Brassard and Gambs \cite{aimeur06} introduce the idea of using the overlap or fidelity $\left|\braket{a}{b}\right|$ of two quantum states $\ket{a}$ and $\ket{b}$ as a `similarity measure'. The fidelity can be obtained through a simple quantum routine sometimes referred to as a \textit{swap test} \cite{buhrman01} (see Figure \ref{figure5}). Given a quantum state $\ket{a,b,0_ {\mathrm{anc}}}$ containing the two wavefunctions as well as an ancilla register initially set to $0$, a Hadamard transformation sets the ancilla into a superposition $\sfrac{1}{\sqrt{2}} (\ket{0} + \ket{1})$, followed by a controlled SWAP-gate on $a$ and $b$ which swaps the two states under the condition that the ancilla is in state $\ket{1}$. A second Hadamard gate on the ancilla results in state $\ket{\psi_{SW}}= \frac{1}{2} \ket{0} (\ket{a, b} + \ket{b, a}) + \frac{1}{2} \ket{1} (\ket{a, b} - \ket{b, a})$ for which the probability of measuring the ground state is given by 
\begin{equation} P(\ket{0_ {\mathrm{anc}}}) = \frac{1}{2} + \frac{1}{2} \left|\braket{a}{b}\right|^2. \label{swap} \end{equation} 
A probability of $\sfrac{1}{2}$ consequently shows that the two quantum states $\ket{a}$ and $\ket{b}$ do not overlap at all (in other words, they are orthogonal), while a probability of $1$ indicates that they have maximum overlap.\\

Based on the swap test, Lloyd, Mohseni and Rebentrost \cite{lloyd13} recently proposed a way to retrieve the distance between two real-valued $n$-dimensional vectors $\vec{a}$ and $\vec{b}$ through a quantum measurement. More precisely, the authors calculate the inner product of the ancilla of state $ \ket{\psi} = \frac{1}{\sqrt{2}} (\ket{0, a} +  \ket{1, b})$ with the state $ \ket{\phi} = \frac{1}{\sqrt{Z}} (|\vec{a}| \ket{0} -  |\vec{b}| \ket{1})$ (with $Z = |\vec{a}|^2 + |\vec{b}|^2$), evaluating $\left|\braket{\phi}{\psi}\right|^2$ as part of a swap test. This looks complicated, but is first of all an inexpensive procedure since the states $\ket{\phi}$ and $\ket{\psi}$ can be efficiently prepared \cite{lloyd13}. The trick lies in the clever definition of a quantum state given in Eq. (\ref{lloyddef}), which encodes the classical length of a vector $\vec{x}$ into the scalar product of the quantum state with itself, $\braket{x}{x} = |\vec{x}|^{-1} |\vec{x}| $. With this definition the identity $|\vec{a} - \vec{b}|^2 = Z \left|\braket{\phi}{\psi}\right|^2$ holds true. The classical distance between two vectors $\vec{a}$ and $\vec{b}$ can consequently be retrieved through a simple quantum swap test of carefully constructed states. Lloyd, Mohseni and Rebentrost use this procedure for a quantum version of the nearest-centroid algorithm. With $\vec{a} \equiv \vec{x}$ and $ \vec{b} \equiv \frac{1}{N_c} \sum_p \vec{v}^p$, they propose to calculate the classical distance from the new input to a given centroid, $|\vec{x} - \frac{1}{N_c} \sum_p \vec{v}^p|$, through the above described procedure. The authors claim that even when considering the operations to construct the quantum states involved, this quantum method is more efficient than the polynomial runtime needed to calculate the same value on a classical computer. \\

Wiebe, Kapoor and Svore \cite{wiebe14} also use a swap test in order to calculate the inner product of two vectors, which is another distance measure between feature vectors. However, they use an alternative representation of classical information through quantum states. Given $n$-dimensional classical vectors $\vec{a},\vec{b}$ with entries $a_j = |a_j| e^{i\alpha_j}, b_j = |b_j|  e^{i\beta_j}, \: j=1,...,n$ as well as an upper bound $r_{\mathrm{max}}$ for the entries of the training vectors in $\mathcal{T}$ and an upper bound for the number of zeros in a vector $d$ (the sparsity), the idea is to write the parameters into amplitudes of the quantum states $\ket{A} = \frac{1}{\sqrt{d}} \sum_j \ket{j} (\sqrt{1-\frac{ |a_j|^2}{ r_{\mathrm{max}}^2}} e^{-i\alpha_j}   \ket{0} + \frac{a_j}{ r_{\mathrm{max}} } \ket{1} ) \ket{1}$ and $\ket{B} = \frac{1}{\sqrt{d}} \sum_j \ket{j}  \ket{1} (\sqrt{1-\frac{ |b_j|^2}{ r_{\mathrm{max}}^2}} e^{-i\beta_j}   \ket{0} + \frac{b_j}{ r_{\mathrm{max}} } \ket{1} )$ and perform a swap test on $\ket{A}$ and $\ket{B}$. According to Eq. (\ref{swap}), the probability of measuring the swap-test ancilla in the ground state is then $P(\ket{0}_ {\mathrm{anc}}) = \frac{1}{2}+\frac{1}{2} | \frac{1}{d r_{\mathrm{max}}^2} \sum_i a_i  b_i |^2$ and the inner product of  $\vec{a},\vec{b}$ can consequently be evaluated by $ | \sum_i a_i  b_i |^2 = d^2 r_{\mathrm{max}}^4 \left( 2 P(\ket{0}_ {\mathrm{anc}}) -1 \right) $,
which is altogether independent of the dimension $n$ of the vector. The authors in fact claim a quadratic speed-up compared to classical algorithms. In the same contribution, Wiebe, Kapoor and Svore also give a scheme for a (weighted) nearest-centroid algorithm based on the Euclidian distance evaluated by well-known algorithms from the toolbox of quantum information, the amplitude estimation algorithm  \cite{brassard00} and D\"urr and H{\o}yer's  \textit{find\_minimum} subroutine \cite{durr96}.   \\

A full quantum pattern recognition algorithm for binary features was presented by Trugenberger \cite{trugenberger02}. He expands his quantum associative memory circuit  \cite{trugenberger01} for this purpose. At the centre is his subroutine to measure the Hamming distance between two binary quantum states. He constructs a quantum superposition containing all states of the quantum training set, and writes the Hamming distance to the binary input vector $\ket{x} = \ket{x_1...x_n}, x_i=\{0,1\}$ into the amplitude of each training vector state. This is done by the following useful routine based on elementary quantum operations. Given two binary strings $\ket{a_1...a_n}$ and $\ket{b_1...b_n}$ with entries $a_i, b_i \in\{0,1\}$, we construct the initial state $\ket{\psi} = \ket{a_1...a_n, b_1...b_n} \otimes \frac{1}{\sqrt{2}}(\ket{0}+\ket{1}) $, consisting of two registers for the qubits of $a$ and $b$ respectively, as well as an extra $2$-dimensional ancilla register in superposition. The inverse Hamming distance between each qubit of the first and second register, 
\[\bar{d}_k = \left\{   \begin{array}{l l}
		    0, & \quad \text{if} \; \ket{a_k} = \ket{b_k},\\
   		    1, & \quad \text{else,} 
  		\end{array} \right . \]
replaces the respective qubit in the second register. This is done by applying an $\mathrm{XOR}_{a,b}$-gate which overwrites the second entry $b_k$ with $0$ if $a_k=b_k$ and else with $1$, as well as a NOT gate. The result is the state
\[\ket{\psi'}= \ket{a_1...a_n,\bar{d}_1...\bar{d}_n} \otimes  \frac{1}{\sqrt{2}} (\ket{0} + \ket{1}).\]
To write the total Hamming distance $\bar{d}_H(\vec{a},\vec{b})$ first into the phase and then into the amplitude, Trugenberger uses the unitary operator $U = \mathrm{exp}({- i\frac{\pi }{2n} H}) $ with $ H = 1 \otimes \sum_k (\frac{1}{2}(\sigma_z +1) )_{d_k} \otimes \sigma_z$ working on the three registers. Note that this adds a negative sign in case the ancilla qubit is in $\ket{1}$. A Hadamard transformation on the ancilla state, $H_{\mathrm{anc}} = 1 \otimes 1\otimes H$ consequently results in 
\begin{multline*} \ket{\psi''} = \mathrm{cos}\left[ \frac{\pi}{2n} \bar{d}_H(\vec{a},\vec{b}) \right]  \ket{a_1...a_n,\bar{d}_1...\bar{d}_n,0} + \\ +\mathrm{sin}\left[ \frac{\pi}{2n} \bar{d}_H(\vec{a},\vec{b})\right]  \ket{a_1...a_n,\bar{d}_1...\bar{d}_n,1}.\end{multline*}
Measuring the ancilla in $\ket{0}$  leads to a state in which the amplitude scales with the Hamming distance between $\vec{a}$ and $\vec{b}$. Of course, the power of this routine only becomes visible if it is applied to a large superposition of training states in the first register $\ket{a_1,...,a_n} \rightarrow \sum_p \ket{v^p}$. A clever measurement then retrieves the states close to the input state with a high probability. \\

\subsection{Quantum computing for support vector machines}

\begin{figure}[t]
\centering    \includegraphics[width=0.3\textwidth]{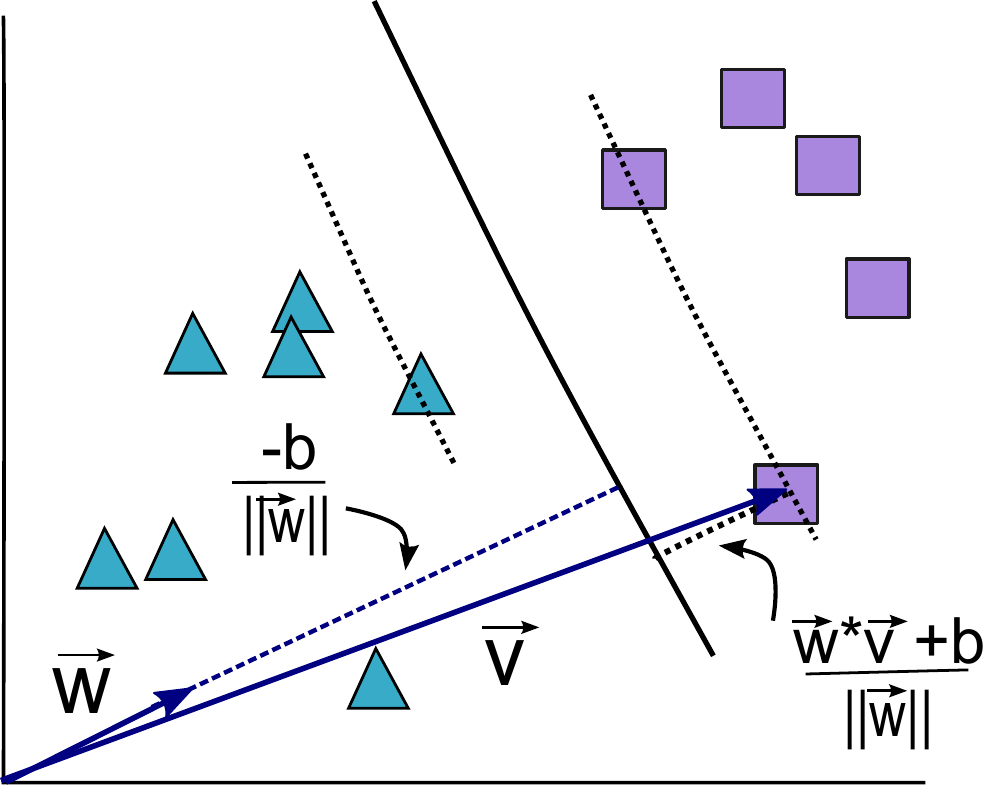} 
\caption{A support vector machine finds a hyperplane (here a line) with maximum margin to the closest vectors. This image illustrates the geometry of the optimisation problem based on \cite{bishop06}. } 
\label{figure6}
\end{figure}

A support vector machine is used for \textit{linear discrimination}, which is a subcategory of pattern classification. The task in linear discrimination problems is to find a hyperplane that is the best discrimination between two class regions and serves as a decision boundary for future classification tasks. In a trivial example of one-dimensional data and only two classes, we would ask which point $x$ lies exactly between the members of class $1$ and $2$, so that all values left of $x$ belong to one class and all values right of $x$ to the other. In higher dimensions, the boundary is given by a hyperplane (see Figure \ref{figure6} for two dimensions). It seems like a severe restriction that methods of linear discrimination require the problem to be linearly separable, which means that there is a hyperplane that divides the datapoints so that all vectors of either class are on one side of the hyperplane (in other words, the regions of each class have to be disjunct). However, a non-separable problem can be mapped onto a linearly separable problem by increasing the dimensions \cite{alpaydin04}.\\

A support vector machine tries to find the optimal separating hyperplane. The best discriminating hyperplane has a maximum distance to the closest datapoints, the so called support vectors. This is a mathematical optimisation problem of finding the maximum margin $\left|\vec{w}\right|^{-1}  (\vec{v}\vec{w} + b)$ between the hyperplane and the support vectors \cite{bishop06} (see Figure \ref{figure6}). In the $2$-dimensional case, the boundary conditions are 
\begin{equation} \begin{aligned}\vec{w}\vec{v}_i + b &\geq 1, \; \mathrm{when}\; c_i = 1, \\ 
\vec{w}\vec{v}_i + b &\leq -1, \; \mathrm{when} \;c_i = -1, \label{SVM}\end{aligned} \end{equation}
for each support vector $\vec{v}_i$ from the training data set and its classification $c_i \in\{-1,1\}$. This means that while finding a maximum margin,  the hyperplane must still separate the training vectors of the two classes correctly. This optimisation problem can be formulated using the Langrangian method \cite{alpaydin04} or in dual space \cite{boser92}. \\

Without going into the complex mathematical details of support vector machines, it is important to note that the mathematical formulation of the optimisation problem contains a kernel $K$, a matrix containing the inner product of the feature vectors $(K)_{pk}= \vec{v}_p \cdot \vec{v}_k,\; p,k=1,...,N$ (or the basis vectors they are composed of) as entries. Support vector machines are in fact part of a larger class of so called kernel methods \cite{bishop06} (for more details see \cite{alpaydin04}) that suffer from the fact that calculating kernels can get very expensive in terms of computational resources. More precisely, quadratic programming problems of this form have a complexity of $O((Nn)^3)$ \cite{bishop06} where $Nn$ is the number of variables involved, and computational resources therefore grow significantly with the size of the training data. It is thus crucial for support vector machines to find a method of evaluating an inner product efficiently. This is where quantum computing comes into play. \\

Rebentrost, Mohseni and Lloyd \cite{rebentrost13} claim that in general, the evaluation of an inner product can be done faster on a quantum computer. Given the quantum state\footnote{The initial state can be constructed by using a Quantum Random Access Memory oracle described in \cite{giovannetti08}, accessing a superposition of memory states in $O(\mathrm{log(nM))}$.} $ \ket{\chi} =  \sfrac{1}{\sqrt{N_{\chi}}} \sum_{i=1}^{2^n} |\vec{x}_i|\ket{i}\ket{x^i} $,
with $N_{\chi} = \sum_{i=1}^{2^n} |\vec{x}^i|^2$. The $\ket{x^i}$ are a $2^n$-dimensional basis of the training vector space $\mathcal{T}$, so that every training vector $\ket{v^p}$ can be represented as a superposition $\ket{v^p} =\sum \alpha_i \ket{x^i}$. Similar to the same authors' distance measurement given in Eq. (\ref{lloyddef}), the quantum evaluation of a classical inner product relies on the fact that the quantum states are normalised as
\[ \braket{x^i}{x^j} = \frac{\vec{x}^i \cdot \vec{x}^j}{|\vec{x}^i| |\vec{x}^j|}. \]
The kernel matrix of the inner products of the basis vectors, $K$ with $(K)_{i,j}= \vec{x}^i \cdot \vec{x}^j$, can then be calculated by taking the partial trace of the corresponding density matrix $\ketbra{\chi}{\chi}$ over the states $\ket{x^i}$,
\[\mathrm{tr_x}[\ketbra{\chi}{\chi}] = \frac{1}{N_{\chi}}\sum \limits_{i,j=1}^{2^n} \underbrace{\braket{x^i}{x^j}|\vec{x}^i||\vec{x}^j|}_{\vec{x}^i \cdot \vec{x}^j} \ketbra{i}{j}  =\frac{\hat{K}}{\mathrm{tr}[K]}\;.\]
Rebentrost, Mohseni and Lloyd propose that the inner product evaluation can not only be used for the kernel matrix but also when a pattern has to be classified, which invokes the evaluation of the inner product between the above parameter vector $\vec{w}$ and the new input (see Eq. \ref{SVM}).\footnote{In the same paper, Rebentrost, Mohseni and Lloyd \cite{rebentrost13} also present another quantum support vector machine that uses the reformulation of the optimisation as a least-squares problem, which appears to be a system of linear equations. Following \cite{harrow09}, this can be solved by a quantum matrix inversion algorithm, which under some conditions (depending on the matrix and the output information required) can be more efficient than classical methods. The classification is then proposed to be done through a swap test.}\\
 
\subsection{Quantum algorithms for clustering}

Clustering describes the task of dividing a set of unclassified feature vectors into $k$ subsets or clusters. It is the most prominent problem in unsupervised learning, which does not use training sets or `prior examples' for generalisation, but rather extracts information on structural characteristics of a data set. Clustering is usually based on a distance measure such as the squared Euclidean distance ($(\vec{a} -\vec{b})^2 \; \mathrm{with} \; \vec{a},\vec{b} \in \mathbb{R}^N$). \\

\begin{figure}[t] 
 \centering    \includegraphics[width=0.21\textwidth]{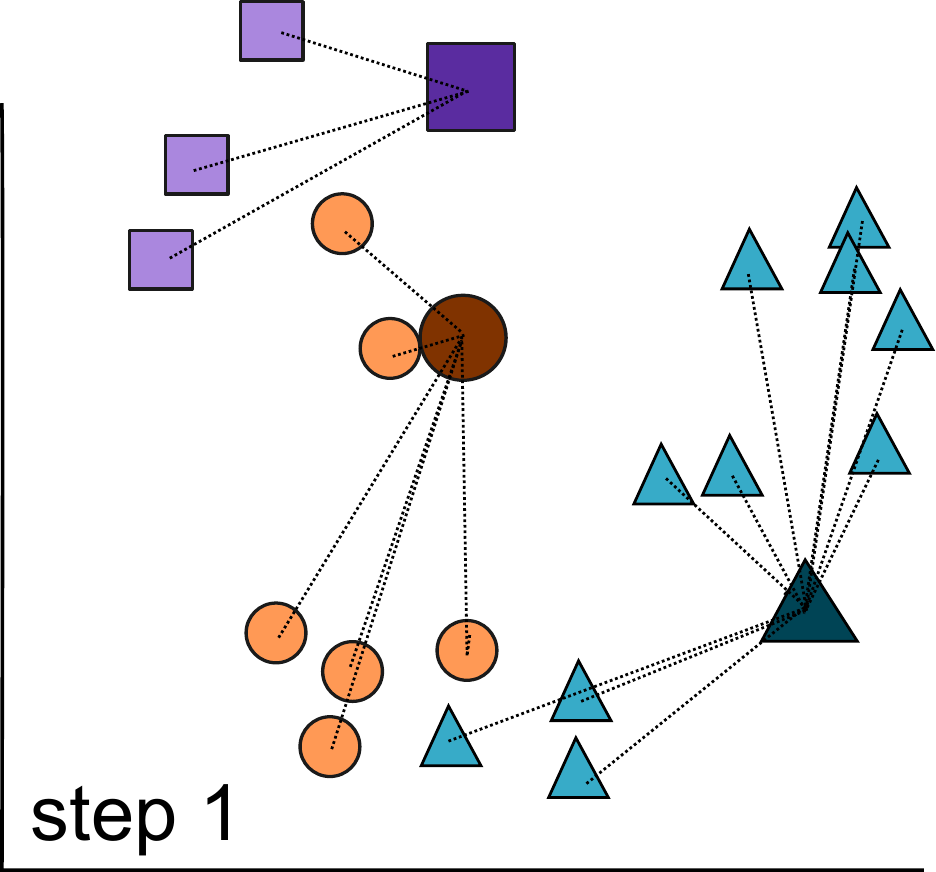} ~~~\includegraphics[width=0.21\textwidth]{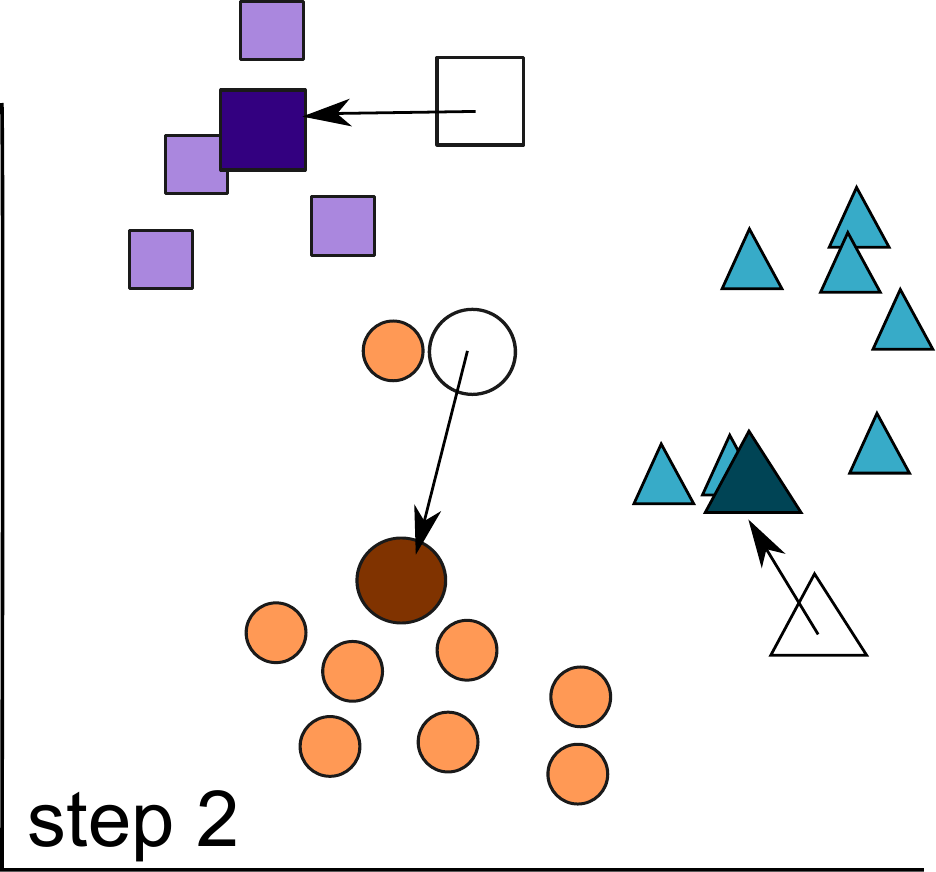} 
\caption{The alternating steps of a $k$-means algorithm. Step 1: The clusters (different shapes and colours) are defined by attributing each vector to the closest centroid vector (larger and darker shapes). Step 2: The centroids of each cluster defined in the previous cycle are recalculated and define a new clustering. } 
\label{figure7}
\end{figure}

The standard textbook example for clustering is the $k$-means algorithm, in which alternately each feature vector or datapoint is assigned to its closest current centroid vector to form a cluster for each centroid, and the centroid vectors get calculated from the clusters of the previous step (see Figure \ref{figure7}). Of course, the first iteration requires initial choices for the centroid vectors, and a free parameter is the number $k$ of clusters to be formed. The procedure eventually converges to stable centroid positions. However, these may represent local minima, as only the position of the initial centroids defines whether a global minima can be reached \cite{rogers12}. Other problems of $k$-means clustering are how to choose the parameter $k$ without prior knowledge of the data, and how to deal with clusters that are visibly not grouped according to distance measures (such as concentric circles). Still, $k$-means works well for many simple applications of reducing many datapoints into only a few groups, for example in data compression tasks. A variation of the $k$-means algorithm is the $k$-median clustering, in which the role of the centroid is taken over by the datapoint of a cluster, that has the smallest total distance to all other points. \\

Besides versions of quantum clustering that are merely \textit{inspired} by quantum mechanics \cite{horn02} or use the quantum mechanical fidelity $\mathrm{Fid}(\ket{\psi},\ket{\phi}) = \left|\braket{\psi}{\phi}\right|^2$ as a distance measure for an otherwise classical algorithm \cite{aimeur06}, several full quantum routines for clustering have been proposed. For example, A{\"\i}meur, Brassard, Gilles and Gambs \cite{aimeur07} use two subroutines for a quantum $k$-median algorithm. First, with the help of an oracle that calculates the distance between two quantum states, the total distance of each state to all other states of one cluster is calculated. Based on the \textit{find\_minimum} subroutine in \cite{durr96}, the authors then describe a routine to find the smallest value of this distance function and select the according quantum state as the new median for the cluster. Unfortunately, the oracle is not described in detail, and their quantum machine learning proposal largely depends on how and with what resources it can be implemented.\\

In their contribution discussed earlier, Lloyd, Mohseni and Rebentrost \cite{lloyd13} present an unsupervised quantum learning algorithm for $k$-means clustering that is based on adiabatic quantum computing. Adiabatic quantum computing is an alternative to the above introduced method of implementing unitary gates, and tries to continuously adjust the quantum system's parameters in an adiabatic process in order to transfer a ground state which is easy to prepare into a ground state which encodes the result of the computation. Although not in focus here, quantum adiabatic computing seems to be an interesting candidate for quantum machine learning methods \cite{pudenz13}. This is why we want to sketch the idea of how to use adiabatic quantum computing for $k$-means clustering.\\

In \cite{lloyd13}, the goal of each clustering step is to have an output quantum superposition $\ket{\chi} = \sfrac{1}{\sqrt{N_c}} \sum_{c, p\in c} \ket{c}\ket{\vec{v}^p}$,
where as usual $\{\ket{v^p}\}_{p=1,...,N}$ is the set of $N$ feature vectors or datapoints expressed as quantum states, and $\ket{c}$ is the cluster the subset  $\{\ket{v^j}\}_{j=1,...,N_c}$ is assigned to after the clustering step. The authors essentially propose to adiabatically transform an initial Hamiltonian $H_0 = 1-\frac{1}{k}\sum_{c,c'} \ketbra{c}{c'}$,
into a Hamiltonian 
\[ H_1 = \sum\limits_{c', j} |\vec{v}^p-\vec{\bar{v}}_{c'}|^2 \ketbra{c'}{c'} \otimes \ketbra{j}{j},\]
encoding the distance between vector $\vec{v}^p$ to the centroid of the closest cluster, $\vec{\bar{v}}_{c}$. They give a more refined version and also mention that the adiabatic method can be applied to solve the optimisation problem of finding good initial or `seed' centroid vectors. \bigbreak

\subsection{Searching for a quantum neural network model}

An artificial neural network is a $n$-dimensional graph where the nodes $x_m$ are called neurons and their connections are weighted by parameters $w_{ml}$ representing synaptic strengths between neurons ($m,l = 1,...,n$). An activation function defines the value of a neuron depending on the current value of all other neurons weighted by the parameters $w_{ml}$, and the dynamics of the neural network is given by successively updating the value of neurons through the activation function. An artificial neural network can thus be understood as a computational device, the input being the initial values of the neurons and the output either a stable state of the entire network or the state of a specific subset of neurons. `Programming' a neural network can be done by selecting weight parameters $w_{ml}$ and an activation function encoding a certain input-output relation. The power of artificial neural networks lies in the fact that they can learn their weights from training data, a fact that neuroscientists believe is the basic principle of how our brain processes information \cite{dayan01}.  \\

\begin{figure}[t] 
  \centering   \includegraphics[width=0.25\textwidth]{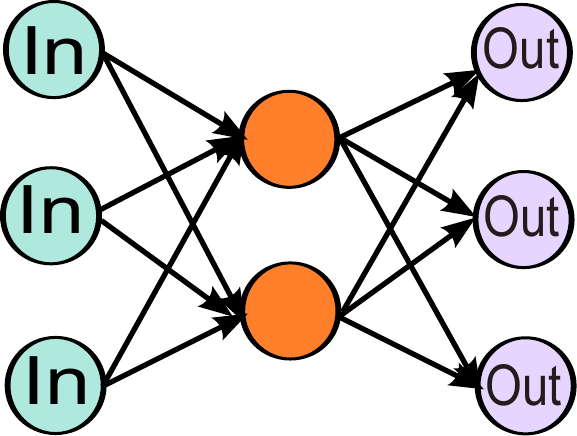}
 \caption{Illustration of a feed-forward neural network with a sigmoid activation function for each neuron. }
 \label{figure8}
\end{figure}

For pattern classification we usually consider so called feed-forward neural networks in which neurons are arranged in layers, and each layer feeds its values into the next layer. An input is presented to a feed-forward neural network by initialising the input layer, and after each layer successively updates its nodes the output (for example encoding the classification of the input) can be read out in the last layer (see Figure \ref{figure8}). \\

Feed-forward neural networks often use sigmoid activation functions    
\[ x_l =  \mathrm{sgm}\left(\sum\limits_{m=1}^N w_{ml} x_m; \kappa \right),\]
defined by $\mathrm{sgm}(a; \kappa) = (1+ \mathrm{e}^{-\kappa a})^{-1}$. If an appropriate set of weight parameters is given, feed-forward neural networks are able to classify input patterns extremely well. To evoke the desired generalisation, the network is initialised with training vectors, the output is compared to the correct output, and the weights adjusted through gradient descent in order to minimise the classification error. The procedure is called backpropagation \cite{hertz91}. A challenge for pattern classification with neural networks is the computational cost for the backpropagation algorithm, even when we consider improved training methods such as deep learning \cite{hinton06}.  \\

There are a number of proposals for quantum versions of neural networks. However, most of them consider another class, so called Hopfield networks, which are powerful for the related task of associative memory that is derived from neuroscience rather than machine learning. A large share of the literature on quantum neural networks tries to find specific quantum circuits that integrate the mechanisms of neural networks in some way \cite{gupta01,oliveira08,silva12,panella11}, trying to use the power of neural computing for quantum computation. A practical implementation is given by Elizabeth Behrman \cite{behrman99,toth00,faber02} who uses interacting quantum dots to simulate neural networks with quantum systems. An interesting approach is also to use fuzzy feed-forward neural networks inspired by quantum mechanics \cite{purushothaman97} to allow for multi-state neurons. Also worth mentioning is the pattern recognition scheme implemented through adiabatic computing with liquid-state nuclear magnetic resonance \cite{neigovzen09}. Despite this rich body of ideas, there is no quantum neural network proposal that delivers a fully functioning efficient quantum pattern classification method that the authors know of. However, it is an interesting open challenge to translate the nonlinear activation function into a meaningful quantum mechanical framework \cite{schuld14b}, or to find learning schemes based on quantum superposition and parallelism. 

\subsection{Towards a quantum decision tree}

\begin{figure}[t] 
 \centering    \includegraphics[width=0.4\textwidth]{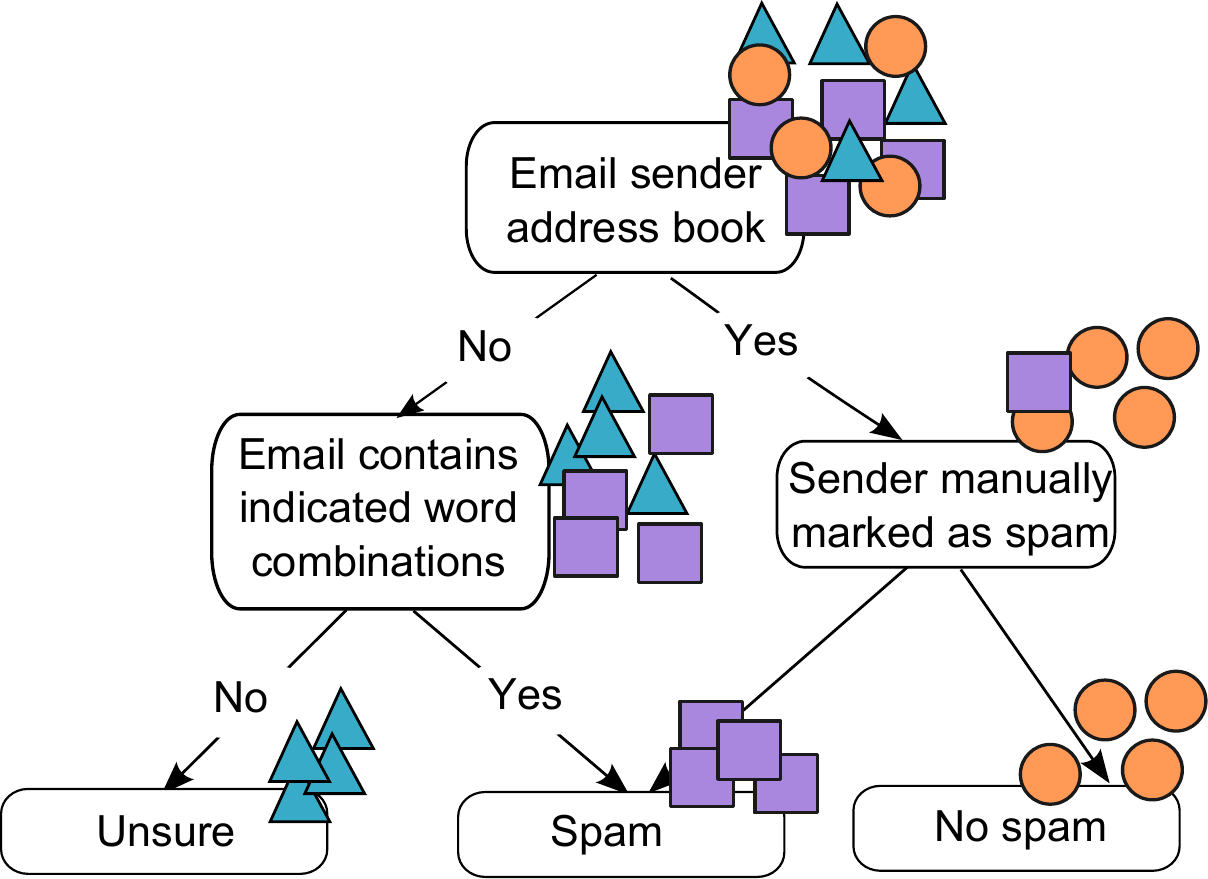} 
\caption{A simple example of a decision tree for the classification of emails. The geometric shapes symbolise feature vectors from different classes that are devided according to decision functions along the tree structure. } 
\label{figure9}
\end{figure}

Decision trees are classifiers that are probably the most intuitive for humans. Depending on the answer to a question on the features, one follows a certain branch leading to the next question until the final class is found (see Figure \ref{figure9}). More precisely, a mathematical tree is an undirected graph in which any two nodes are connected by exactly one edge. Decision trees in particular have one starting node, the `root' (a node with outgoing but no incoming edges), and several end points or `leaves' (nodes with incoming but no outgoing edges). Each node except from the leaves contains a decision function which decides which branch an input vector follows to the next layer, or in other words, which partition on a set of data is makes. The leaves then represent the final classification. As in the example in Figure \ref{figure9}, this procedure could be used to classify an email as `spam', `no spam' or `unsure'. \\

Decision trees, as all classifiers in machine learning, are constructed using a training data set of feature vectors. The art of decision tree design lies in the selection of the decision function in each node. The most popular method is to find the function that splits the given dataset into the `most organised' sub-datasets, and this can be measured in terms of Shannon's entropy \cite{alpaydin04}. Assume the decision function of a node splits a set of $P$ feature vectors $\{\vec{v}^p\}$, $p=1,...,N$ into M subsets each containing $\{N_1,...,N_M\}$ vectors respectively (and $\sum_{i=1}^{M} N_i = N$). Without further information, we calculate the probability of any vector $\vec{v}^p$ to be attributed to subset $i$, $i\in \{1,...,M\}$ (in other words to proceed to the $i$th node of the next layer) as $ \rho^i = \frac{N
_i}{N}$, and the entropy caused by the decision function or partition is consequently $S = -\sum_{i=1}^M \rho^i \mathrm{log}(\rho^i)$. For example, in a binary tree where all nodes have two outgoing edges, the best partition would split the original set into two subsets of the same size. Obviously, this is only possible if one of the features allows for such a split. Depending on the application, an optimal decision tree would be small in the number of nodes, branches and/or levels. \\

Lu and Brainstein \cite{lu14} propose a quantum version of the decision tree. Their classifying process follows the classical algorithm with the only difference that we use quantum feature states $\ket{v}^p = \ket{v_{1}^p,...,v_{n}^p}$ encoding $n$ features into the states of a quantum system. At each node of the tree, the set of training quantum states is divided into subsets by a measurement (or as the authors call it, estimating attribute $v_i,\; i = 1,...,n$). Lu and Brainstein do not give a clear account of how the division of the set at each node takes place and remain enigmatic in this essential part of the classifying algorithm. They contribute the interesting idea of using the von Neumann entropy to design the graph partition. Although the first step has been made, the potential of a quantum decision tree is still to be established.

\subsection{Quantum state classification with Bayesian methods}

Stochastic methods such as Bayesian decision theory play an important role in the discipline of machine learning. It can also be used for pattern classification. The idea is to analyse existing information (represented by the above training data set $\mathcal{T}$) in order to calculate the probability that a new input is of a certain class. An illustrative example is the risk class evaluation of a new customer to a bank. This is nothing else than a conditional probability and can be calculated using the famous Bayes formula
\[p(c|\vec{x}) = \frac{p(c)p(\vec{x}|c)}{p(\vec{x})}.\] 
Here, $p(c), p(\vec{x})$ are the probabilities of data being in class $c$ and of getting input $\vec{x}$ respectively, while $p(c|\vec{x})$ is the conditional probability of assigning $c$ upon getting $\vec{x}$ and $p(\vec{x}|c)$ is the class likelihood of getting $\vec{x}$ if we look in class $c$. Obviously, we assign the class with the highest conditional probability (or `Bayes classifier') $p(c_l|\vec{x})$ to an input \cite{alpaydin04}. Values of interest, such as risk functions, can be calculated accordingly. Bayesian theory is an interesting candidate for the translation into quantum physics, since both approaches are probabilistic. \\

Opposed to above efforts to improve machine learning algorithms through quantum computing, Bayesian methods can be used for an important task in quantum information called quantum state classification. This problem stems from quantum information theory itself, and the goal is to use machine learning based on Bayesian theory in order to discriminate between two quantum states produced by an unknown or partly unknown source. This is again a classification problem, since we have to learn the discrimination function between two classes $c_1, c_2$ from examples. The two (unknown) quantum states are represented by density matrices $\rho, \sigma$. The basic idea is to use a positive operator-valued measurement (POVM) with binary outcome corresponding to the two classes as a Bayesian classifier, in other words, to learn (or calculate) the measurement on our quantum states that is able to discriminate them \cite{gutja10}. For this process we have a training set consisting of examples of the two states and their respective classification, $\mathcal{T} = \{(\rho, c_1), (\sigma, c_2), (\rho, c_1), ...\}$ and the experimenter is allowed to perform any operation on the training set. Gu{\c{t}}{\u{a}} and Kot{\l}owski \cite{gutja10} find an optimal qubit classification strategy while Sasaki and Carlini \cite{sasaki01} are concerned with the related template matching problem\footnote{Template matching is the task to assign the most similar training vector of a training set to an input vector.} by solving an optimisation problem for the measurement operator. Sentis et al. \cite{sentis12} give a variation in which the training data can be stored as classical information. The proposals are so far of theoretical nature and await experimental verification of the usefulness of this scheme.

\subsection{Hidden quantum Markov models}

\begin{figure}[t]
 \centering    \includegraphics[width=0.3\textwidth]{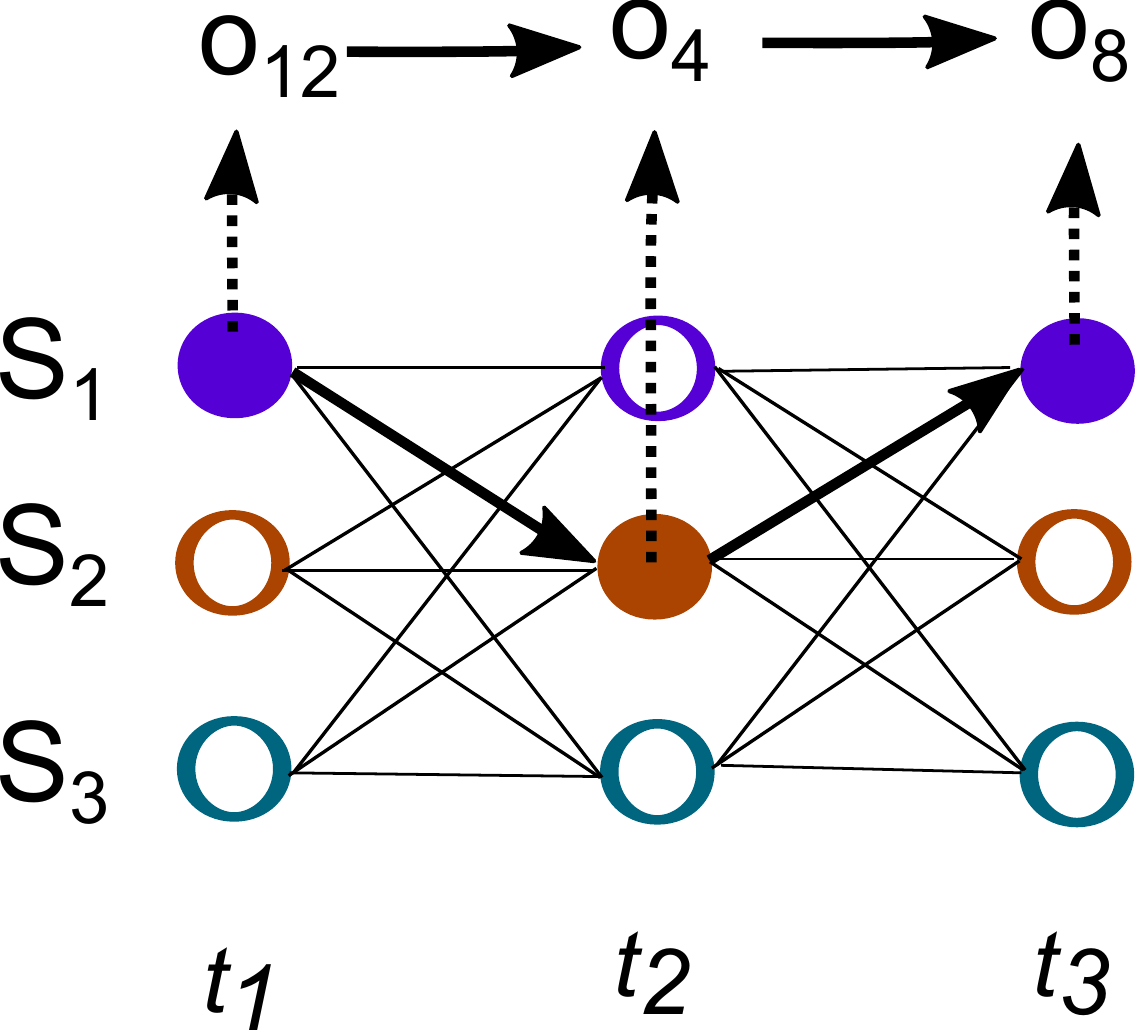} 
\caption{(Colour online) A hidden Markov model is a stochastic process of state transitions. In this sketch, the three states $s_1,s_2,s_3$ are connected with lines symbolising transition probabilities. A deterministic realisation is a sequence of states, here the transition $s_1 \rightarrow s_2 \rightarrow s_1$ that give rise to observations $o_{12} \rightarrow o_4 \rightarrow o_8$. A task for hidden Markov models is to guess the most likely state sequence given an observation sequence.    } 
\label{figure10}
\end{figure}

In the last couple of years, hidden Markov models were another important method of machine learning that has been investigated from the perspective of quantum information \cite{monras10, clark14}. Hidden Markov models are Markov processes for which the states of the system are only accessible through observations (see Figure \ref{figure10}, for a very readable introduction see \cite{rabbiner89}). In a (first order discrete and static) Markov model, a system has a countable set of states $\mathcal{S}=\{ s_m\}_{m=1,...,M}$ and the transition between these states are governed by a stochastic process in such a way that given a set of transition probabilities $\{a_{ml}\}_{m,l = 1,...,M}$, the system's state at time $t+1$ only depends on the previous state at time $t$. In a hidden model, the state of the system is only accessible through observations at time $t$ $\{ o_t\}$ that can take one of a set of symbols, and an observation again has a certain probability to be invoked by a specific state. Hidden Markov models are thus doubly embedded stochastic processes. To use a common application for pattern recognition as an example \cite{bishop06}, consider a recorded speech. The speech is a realisation of a Markov process, a so called Markov chain of successive words. The recording is the observation, and we shall for now imagine a way to translate the signal into discrete symbols. A Markov model is defined by the transition probabilities between words in a certain language, and the model can be learned from examples of speeches. A hidden Markov model also includes the conditional probabilities that given a certain signal observation, a certain word has been said. Goals of such models are to find the sequence of words that is the most likely for a recording, to predict the next word or, if only given the recording, to infer the optimal hidden Markov model that would encode it. Hidden Markov models play an important role in many other applications such as DNA analysis and online handwriting recognition \cite{bishop06}.\\

Monras, Beige and Wiesner \cite{monras10} first introduced a hidden quantum Markov model in 2010. In contrast to a previous paper \cite{wiesner08} in which the observations are represented by quantum basis states and the observation process is given by a von Neumann or projective measurement of an evolving quantum system, the authors consider the much more general formalism of open quantum systems (for an introduction to open quantum systems, see \cite{breuer02}). The state of a system is given by a density matrix $\rho$ and transitions between states are governed by completely positive trace-nonincreasing superoperators $\mathcal{A}^i$ acting on these matrices. These operations can always be represented by a set of Kraus operators \cite{breuer02} $\{\mathcal{K}^i_1,..., \mathcal{K}^i_q\}$ fulfilling the probability conservation condition $\sum_q \mathcal{K}_q^{i \dagger} \mathcal{K}_q^i \leq 1$,
\[\rho' = \mathcal{A}^i \rho = \sum_k \mathcal{K}^i_k \rho\mathcal{K}^{i \dagger}_k.\]
The probability of obtaining state $\rho_{s} = P(\rho_{s})^{-1} \mathcal{A}^s \rho$ is given by $P(\rho_{s}) = \textrm{tr}[\mathcal{A}^s \rho]$ \cite{monras10}. \\

The advantage of hidden quantum Markov models is that they contain classical hidden Markov models and are therefore a generalisation offering richer dynamics than the original process \cite{monras10}. In future there might also be the possibility of `calculating' the outcomes of classical models via quantum simulation. That would be especially interesting if the quantum setting could learn models from given examples, a problem which is nontrivial \cite{rabbiner89}. Clark et al. \cite{clark14} add the notion that hidden quantum Markov models can be implemented using open quantum systems with instantaneous feedback, in which information obtained from the environment  is used to influence the system. However, a rigorous treatment of this idea is still outstanding, and the power of hidden quantum Markov models to solve the problems for which classical models where developed is yet to be shown.\\

An interesting sibling of  hidden quantum Markov models are quantum observable Markov decision processes \cite{barry14} which use a very similar idea. Classical observable Markov decision processes can be understood as hidden Markov models in which before each step an agent takes a decision for a certain action, leading to the next state of the system. The state of the system is again only accessible through observations that deliver probabilistic information. The goal is to find a strategy (defining what action to take upon what observation) that maximises the rewards given by a reward function. This is a problem of reinforcement learning by intelligent agents which is not the focus of this contribution. However, we also find the striking analogy to Kraus operations on open quantum systems representing the actions that manipulate the density matrix or stochastic description of the system.

\section{Conclusion}\label{conc}
This introduction into quantum machine learning gave an overview of existing ideas and approaches to quantum machine learning. Our focus was thereby on supervised and unsupervised methods for pattern classification and clustering tasks, and it is therefore by no means a complete review. In summary, there are  two main approaches to quantum machine learning. Many authors try to find quantum algorithms that can take the place of classical machine learning algorithms to solve a problem, and show how an improvement in terms of complexity can be gained. This is dominantly true for nearest neighbour, kernel and clustering methods in which expensive distance calculations are sped up by quantum computation. Another approach is to use the probabilistic description of quantum theory in order to describe stochastic processes. In the case of hidden quantum Markov models, this served to generalise the model, while Bayesian theory was also used for genuinely quantum information tasks like quantum state discrimination. A great deal of contributions is still in a phase of exploring possibilities to combine formalisms from quantum theory and methods of machine learning, as seen in the area of quantum neural networks and quantum decision trees.\\

As previously remarked, a quantum theory of learning is yet outstanding. Although working on quantum machine learning algorithms, only very few contributions actually answer the question of how the strength and defining feature of machine learning, the \textit{learning} process, can actually be simulated in quantum systems. Especially learning methods of parameter optimisation have not yet been accessed from a quantum perspective. Different approaches to quantum computing can be investigated for this purpose. In quantum computing based on unitary quantum gates, the challenge would be to parameterise and gradually adapt the unitary transformations that define the algorithm. Several ideas in that direction have been investigated already  \cite{gammelmark09, gammelmark13, bisio10}, and important tools could be quantum feedback control \cite{hentschel10} or quantum Hamiltonian learning \cite{wiebe14b}. As mentioned before, adiabatic quantum computing might lend itself to learning as an optimisation problem \cite{pudenz13}. Other alternatives of quantum computation, such as dissipative \cite{verstraete09} and measurement-based quantum computing \cite{briegel09} might also offer an interesting framowork for quantum learning. In summary, even though there is still a lot of work to do, quantum machine learning remains a very promising emerging field of research with many potential applications and a great theoretical variety.

\section*{Acknowledgements} 

This work is based upon research supported by the South African Research Chair Initiative of the Department of Science and Technology and National Research Foundation.

\end{document}